\documentclass{article}
\usepackage{graphicx} 
\usepackage{geometry} 
\usepackage{fancyhdr} 
\usepackage{amsmath,amsfonts,amsthm}
\usepackage{amssymb}
\usepackage{setspace}
\usepackage{parskip}
\usepackage{booktabs}
\usepackage{tikz}

\nocite{*}

\title{Some MDS codes over dihedral groups}
\author{YuChao Wang}
\date{October 2024}

\begin{document}

\maketitle

\hrule
\section*{Abstract} 
In this paper,we show some $[2n,2n-2,3]$ and $[2n,2n-3,4]$ MDS codes over dihedral codes $F_qD_{2n}$\,,\,in the case $n$ is odd and  char$F_q$$\nmid$$\lvert G \rvert$ and $F_q$ contains primitive root of exponent $\lvert G \rvert$ i.e $F_q$ is the splitting field of $G$\,.\,Before that,we will give the Wedderburn decomposition and specific forms of linear primitive idempotents of $F_qD_{2n}$ under the above conditions.The MDS codes we construct are obtained by its Wedderburn decomposition and linear primitive idempotents.
\\
\hrule
\hspace*{\fill} \\
Keywords\,:\,Group algebra, diheral group, MDS codes, central idempotents, Wedderburn decomposition

\section{Introduction}
Let G be a finite group , $\mathbb{F}$ a field , $\mathbb{F}G:=\{\sum_{g\in{G}}a_gg|a_g\in{\mathbb{F}}\}$ denotes group algebra . In particular , $\mathbb{F}G$ is an $\mathbb{F}$-vector space with the basis $\{g\in{G}\}$ , equipped with addition and multiplication as follow
\begin{equation*}
  (\sum_{g\in{G}}a_gg)+(\sum_{g\in{G}}b_gg)=\sum_{g\in{G}}(a_g+b_g)g
\end{equation*}
\begin{equation*}
  (\sum_{g\in{G}}a_gg)(\sum_{g\in{G}}b_gg)=\sum_{g\in{G}}(\sum_{h\in{G}}a_hb_{h^{-1}g})g
\end{equation*}\\
Let $F_q$ be a finite field of order $q$ , any left-ideal I of $F_qG$ is call a group code or $G$-code . Let $u={\sum\limits_{g\in{G}}a_gg}$ $\in{F_qG}$ , the weight of $u$ is defined as $wt(u)$:=$|\{g\in{G}|a_g\neq{0}\}|$ . Generally , we consider C as a $G$-code . The paremeter of $C$ is $[n,k,d]$ , where $n=|G|$ , $k$=dim$_{F_q}(C)$ , $d=d(C)$:=$\mathop{min}\limits_{c\in{C},c\neq{0}}w(c)$ . The dihedral group
\begin{align*}
   D_{2n}:=< a,b|a^{n}=1,b^{2}=1,ab=ba^{-1}>
   =\{e,a,a^{2},\cdots,a^{n-1},b,ba,ba^{2},\cdots,ba^{n-1}\}
   \end{align*}
Now we define the mapping between $F_qD_{2n}$ and $F_q^{2n}$ as follow . Let $F_{_{2n}}=\{f_{_1},\cdots,f_{_{2n}}\}$ be a standard basis of $F_{q}^{2n}$ . Let $\phi:D_{2n}\rightarrow F_q^{2n}$ be the bijective map , given by 
\begin{equation}
    \phi(a^{i})=f_{_{i}},\qquad\phi(ba^{i})=f_{_{n+i+1}}\quad(i=0,\cdots,n-1)
    \end{equation}
The map $\phi$ can naturally be extended to the linear isomorphism $\phi:F_qD_{2n}\rightarrow F_q^{2n}$ . Let $I\subset F_qD_{2n}$ be a dihedral code , and let $S=\{s_{_1},\cdots,s_{_k}\}$ be a basis of it . The matrix 
\begin{equation*}
    G_{I}=\begin{pmatrix}
        \rule[1pt]{0.5cm}{0.1em}\phi(s_{_1})\rule[1pt]{0.5cm}{0.1em}\\
        \rule[1pt]{0.5cm}{0.1em}\phi(s_{_2})\rule[1pt]{0.5cm}{0.1em}\\
        \vdots\\
        \rule[1pt]{0.5cm}{0.1em}\phi(s_{_k})\rule[1pt]{0.5cm}{0.1em}
    \end{pmatrix}
\end{equation*}
is called a generating matrix of $I$ .\\
From \cite{bib16}\cite{bib12} , there is one to one correspondence between left-ideal and right-ideal by anti-automorphisim $\ast$ , where $\ast$ : $F_qG$$\rightarrow$$F_qG$ is defined as follow
\begin{equation*}
   \ast: {\sum_{g\in{G}}a_gg}\longmapsto{\sum_{g\in{G}}a_gg^{-1}}
\end{equation*}
The auti-automorphism $\ast$ keep the distance and dimension of one-sided ideals invariant . That is we just need to study one-sided ideal , and we consider left-ideal throughout this paper.\\
In \cite{bib4} , the constructure of primitive idempotents of $G$ if it is Abelian and $F_q$ contains all the $m-th$ roots of unity where $m$ is the exponent of $G$ . In \cite{bib17} , the author gave the strcture of all representation of dihedral group $D_{2n}$ over the complex field  $\mathbb{C}$ . In \cite{bib7} , the explicit mapping from $F_qD_{2n}$ to its Wedderburn decomposition was given by F.E. Brochero Martínez when gcd($2n,q$)=1 . Kirill V. Vedenev and  Vladimir M. Deundyak gave its inverse mapping , simultaneously generalized the results to the case gcd($n,q$)=1 in \cite{bib2} . Since the mapping from $F_qD_{2n}$
 to its Wedderburn decomposition is $F_q$-algebra isomorphism , then we can use the Wedderburn decomposition of $F_qD_{2n}$ to derive its left-ideals . In \cite{bib6}, Sudesh Sehrawat described $[2n,2n-1,2]$ MDS codes and $[2n,2n-2,2]$ group codes of $F_qD_{2n}$ for every $n$ . In this paper,we generalize the result of Sudesh Sehrawat in the case of n is odd , describe some $[2n,2n-2,3]$ and $[2n,2n-3,4]$ MDS codes of $F_qD_{2n}$ . In section 2 , the Wedderburn decomposition of $F_qD_{2n}$ is given when $F_q$ contains all the $m-th$ . In section 3 , we will show the explicit forms of central primitive idempotent of $F_qD_{2n}$ and $F_qC_{n}$ under the condition $F_q$ contains all $2n-th$ roots of unity , where $C_{n}$ is the $n$-cyclic group of $D_{2n}$ . In section 4 , we construct some MDS dihedral codes by its Wedderburn decomposition . In section 5 , we give an example .

\section{The Wedderburn decomposition of $F_qD_{2n}$}
In this section , let $D_{2n}$ be dihedral group , $F_q$ denotes finite field of oreder $q$ such that char$F_q$$\nmid$$\lvert G \rvert$ . For every poylnominal $g(x)$ with $g(0)\neq{0}$ , $g^{*}(x):=x^{deg(g)}g(\dfrac{1}{x})$ denotes the reciprocal polynomial of $g(x)$ . For any polynomial $g$ satisfies that $g$ and $g^{*}$ have the same roots in its splitting field , we say that $g$ is auto-reciprocal polynomial. From \cite{bib7},The polynomial $x^{n}-1\in{F_q[x]}$ can be split into monic irreducible factors as follow:
\begin{equation*}
   x^{n}-1=f_{1}f_{2}\cdots f_{r}f_{r+1}f_{r+1}^{*}f_{r+2}f_{r+2}^{*}\cdots f_{r+s}f_{r+s}^{*}
\end{equation*}
where $f_{1}=x-1$ , $f_{2}=x+2$ if $n$ is even , and $f_{j}=f_{j}^{*} \quad\forall 1\leq j\leq r$ . That is $f_{1},f_{2}\cdots,f_{r}$ are auto-reciprocal polynomials and $f_{r+1},f_{r+1}^{*},f_{r+2},f_{r+2}^{*},\cdots,f_{r+s},f_{r+s}^{*}$ are non-auto-reciprocal polynomials . Let $\alpha_j$ denotes a root of the polynomial $f_j$ . 
Define a number $
\delta = \left\{
\begin{array}{ll}
1 & ,\text{$n$ is odd }  \\
2 & ,\text{$n$ is even} 
\end{array}
\right.
$

\textbf{Theorem 2.1}(\cite{bib7},\textbf{Theorem 3.7})\textbf{.}The group algebra $F_qD_{2n}$ has Wedderburn decomposition of the form 
\begin{equation}
    P=\mathop{\bigoplus}\limits_{j=1}^{r+s}P_j:F_qD_{2n}\simeq\mathop{\bigoplus}\limits_{j=1}^{r+s}A_j
\end{equation}
where
\begin{equation}
    A_j=\begin{cases}
        F_q\oplus F_q, & j\leq\delta  \\
        M_2(F_q[\alpha_j+\alpha_j^{-1}]), & \delta+1\leq j\leq r\\
        M_2(F_q[\alpha_j]), & r+1\leq j\leq r+s.
    \end{cases}
\end{equation}
and\\
\begin{equation}
   P_j=\begin{cases}
   \gamma_j, & 1\leq j\leq \delta\\
   \sigma_j\tau_j, & \delta+1\leq j\leq r\\
   \tau_j, & r+1\leq j\leq r+s
   \end{cases}
\end{equation}
\textbf{Remark}: $\tau_j$ and $\sigma_j$ is given as follow
\begin{align*}   
    \gamma_1:F_qD_{2n}&\rightarrow F_q\oplus F_q\\
    a\quad&\mapsto(1,1)\\
    b\quad&\mapsto(1,-1)
\end{align*}
and
\begin{align*}   
    \gamma_2:F_qD_{2n}&\rightarrow F_q\oplus F_q\\
    a\quad&\mapsto(-1,-1)\\
    b\quad&\mapsto(1,-1)
\end{align*}
and for every $j\geq\delta+1$ ,
\begin{align*}
    \tau_j:F_qD_{2n}&\rightarrow M_2(F_q[\alpha_j])\\
    a\quad&\mapsto\begin{pmatrix}
        \alpha_j&0\\
        0&\alpha_j^{-1}
    \end{pmatrix}\\
    b\quad&\mapsto\begin{pmatrix}
        0&1\\
        1&0
    \end{pmatrix}
\end{align*}
For $\delta+1\leq j\leq r$ consider the automorphism $\sigma_j$ of the algebra $M_2(F_q[\alpha_j])$ given by
\begin{equation*}
    \sigma_j(X):Z_j^{-1}XZ_j\quad,\quad Z_j:=\begin{pmatrix}
        1&-\alpha_j\\
        1&-\alpha_j^{-1}
    \end{pmatrix}
\end{equation*}
\textbf{Corollary 2.2}\quad If $2n|q-1$ , let $\xi$ denotes $n-th$ primitive root of unity , then $x^n-1\in{F_q[x]}$ can split as follow
\begin{equation}
x^n-1=\left\{
\begin{array}{ll}
(x-1)(x-\xi)(x-\xi^{-1})\cdots(x-\xi^{\frac{n-1}{2}})(x-\xi^{-\frac{n-1}{2}}) & ,\text{$n$ is odd }  \\
(x-1)(x-2)(x-\xi)(x-\xi^{-1})\cdots(x-\xi^{\frac{n-2}{2}})(x-\xi^{-\frac{n-2}{2}}) & ,\text{$n$ is even} 
\end{array}
\right.
\end{equation}
\begin{equation}
    P=\mathop{\bigoplus}\limits_{j=1}^{\delta+\frac{n-\delta}{2}}P_j:F_qD_{2n}\simeq\mathop{\bigoplus}\limits_{j=1}^{\delta+\frac{n-\delta}{2}}A_j
\end{equation}
where
\begin{equation}
    A_j=\begin{cases}
        F_q\oplus F_q, & j\leq\delta  \\      
        M_2(F_q), & \delta+1\leq j\leq \frac{n-\delta}{2}.
    \end{cases}
\end{equation}
and\\
\begin{equation}
   P_j=\begin{cases}
   \gamma_j, & 1\leq j\leq \delta\\
   \tau_j, & \delta+1\leq j\leq \frac{n-\delta}{2}
   \end{cases}
\end{equation}
\begin{proof}
    It is well known , $F_q^{*}$ is a cyclic group of order $q-1$ , equipped with multiplication . From $2n|q-1$ , we know that $F_q$ has all $n-th$ roots of unity , then $(x^n-1)$ splits as (4) is obvious . The rest of result follows from \textbf{Theorem 2.1} immediately .
\end{proof}
\textbf{Remark}. From \{\cite{bib3},\textbf{Theorem 1.2}\} , we know that the Wedderburn decomposition of semisimple algebra is unique up to isomorphism . Let $R=F_qD_{2n}$ , if $2n|q-1$ ,  ${End_R(S)^{\circ}}\simeq F$ , that is ${End_R(S)^{\circ}}={End_R(S)}\simeq F$ , for any $S\in{\overline{Irr_{F_q}}D_{2n}}$ , then $F_q$ is the splitting field of $D_{2n}$ .

\section{Idempotents}
In this section , let $A$ be an Abel group , whose order and exponent we denote by $n$ and $m$ . Let $F_q$ be a finite field containing all the $m-th$ roots of unity . Define $G$ be the set of irreducible character of $A$ over $F_q$ . As well known , $A\simeq Z_{m_1}\times Z_{m_2}\times\cdots\times Z_{m_t}$ , where $Z_{m_i}$ is cyclic group of order $m_i$ and $m_1|m_2|\cdots|m_t$ . We identify $Z_{m_1}\times Z_{m_2}\times\cdots\times Z_{m_t}$ with $A$ . To pretend confusion with the multiplication of group algebra , we denote the operation in $Z_{m_1}\times Z_{m_2}\times\cdots\times Z_{m_t}$ by multiplication i.e $(x_1,x_2,\cdots,x_t)\cdot(y_1,y_2\cdots,y_t):=(x_1+y_1,x_2+y_2,\cdots,x_t+y_t)$ . The exponent of $A$ is $m_t=m$ , then $F_q$ has all $m_{i}-th$ roots , denotes $\xi_i$ be $m_i-th$ primitive root of unity , for $\;1\leq i\leq t$ . \\
\quad\\
\textbf{Theorem 3.1}. Let $A$ be an Abel group , whose order and exponent we denote by $n$ and $m$ . Let $F_q$ be a finite field containing all the $m-th$ roots of unity , then the irreducible character of $A$ can be express in the following form 
\begin{equation}
    \chi_{_{(a_1,a_2,\cdots,a_t)}}(x_1,x_2,\cdots,x_t):=\xi_1^{a_1x_1}\xi_2^{a_2x_2}\cdots\xi_t^{a_tx_t}
\end{equation}
where $(x_1,x_2,\cdots,x_t)\in{Z_{m_1}\times Z_{m_2}\times\cdots\times Z_{m_t}}$ , $1\leq a_i\leq m_i$ for all $i$ . 
\begin{proof}
    From [] , $F_q$ is the splitting field of $A$ , the degree of irreducible character of $A$ is 1 , and $A$ exactly has $|A|$ nonisomorphic character . It is enough to prove $\chi_{(a_1,a_2,\cdots,a_t)}$ is irreducible character of $A$ and $deg\chi_{_{(a_1,a_2,\cdots,a_t)}}=1$ . By $\chi_{_{(a_1,a_2,\cdots,a_t)}}(0,0,\cdots,0)=1$ , $\chi_{_{(a_1,a_2,\cdots,a_t)}}$ is integral linear combination of irreducible character . Let $V$ be $F_q$-linear space of dimension 1 , define 
    \begin{equation*}
        \rho_{_{(a_1,a_2,\cdots,a_t)}}(x_1,x_2,\cdots,x_t)v:=\xi_1^{a_1x_1}\xi_2^{a_2x_2}\cdots\xi_t^{a_tx_t}v\qquad ,\forall v\in{V}
    \end{equation*}
    Observe that $(V,\rho_{_{(a_1,a_2,\cdots,a_t)}})$ is $F_q$-representation of $A$ , whose character is $\chi_{_{(a_1,a_2,\cdots,a_t)}}$ exactly . Now we proof $(V,\rho_{_{(a_1,a_2,\cdots,a_t)}})$ is not isomorphic to $(V,\rho_{_{(b_1,b_2,\cdots,b_t)}})$ when $(a_1,a_2,\cdots,a_t)\neq(b_1,b_2,\cdots,b_t)$ . It's equivalent to proof $(\chi_{_{(a_1,a_2,\cdots,a_t)}},\chi_{_{(b_1,b_2,\cdots,b_t)}})=0$ and
    \begin{align*}
        (\chi_{_{(a_1,a_2,\cdots,a_t)}},\chi_{_{(b_1,b_2,\cdots,b_t)}})&=\dfrac{1}{|A|}\sum_{{(x_1,x_2,\cdots,x_t)}\in{A}}\chi_{_{(a_1,a_2,\cdots,a_t)}}(x_1,x_2,\cdots,x_t)\chi_{_{(b_1,b_2,\cdots,b_t)}}(-x_1,-x_2,\cdots,-x_t)\\
        &=\dfrac{1}{|A|}\sum_{{(x_1,x_2,\cdots,x_t)\in{A}}}\xi_1^{(a_1-b_1)x_1}\xi_2^{(a_2-b_2)x_2}\cdots\xi_t^{(a_t-b_t)x_t}
    \end{align*}
From $(a_1,a_2,\cdots,a_t)\neq(b_1,b_2,\cdots,b_t)$ , without loss of generality , let $a_1\neq b_1$ ,
    \begin{equation*}
        \text{Above equation}=\dfrac{1}{|A|}\sum_{(x_2,\cdots,x_t)}(\sum_{x_1}\xi_1^{(a_1-b_1)x_1})\xi_2^{(a_2-b_2)x_2}\cdots\xi_t^{(a_t-b_t)x_t}
    \end{equation*}
Observe that $\sum\limits_{x_1}\xi_1^{(a_1-b_1)x_1}=\xi_1+\xi_1^{(a_1-b_1)}+\cdots+\xi_1^{(a_1-b_1)(m_1-1)}=\dfrac{\xi_1(1-\xi_1^{(a_1-b_1)m_1})}{1-\xi_1^{a_1-b_1}}=0$ , because $\xi_1^{_{(a_1-b_1)}}\neq 1$ and $\xi_{_1}^{\scriptscriptstyle{m}}=1$ , then $(\chi_{_{(a_1,a_2,\cdots,a_t)}},\chi_{_{(b_1,b_2,\cdots,b_t)}})=0$ .  
\end{proof}
Define multiplication in $G$ by 
\begin{align*}
(\chi_{_{(a_1,a_2,\cdots,a_t)}}\cdot\chi_{_{(b_1,b_2,\cdots,b_t)}})(x_1,x_2,\cdots,x_t):&=\chi_{_{(a_1,a_2,\cdots,a_t)}}(x_1,x_2,\cdots,x_t)\cdot\chi_{_{(b_1,b_2,\cdots,b_t)}}(x_1,x_2,\cdots,x_t)\\
&=\xi_1^{(a_1+b_1)x_1}\xi_2^{(a_2+b_2)x_2}\cdots\xi_t^{(a_t+b_t)x_t}\\
&=\chi_{_{(a_1+b_1,a_2+b_2,\cdots,a_t+b_t)}}(x_1,x_2,\cdots,x_t)
\end{align*}
Hence $\chi_{_{(a_1,a_2,\cdots,a_t)}}\cdot\chi_{_{(b_1,b_2,\cdots,b_t)}}=\chi_{_{(a_1+b_1,a_2+b_2,\cdots,a_t+b_t)}}$\\
\quad\\
\textbf{Corollary 3.2}. The set of irreducible character of $A$ denotes by $G$ is isomorphic to $A$ as group.
\begin{proof}
Define mapping
    \begin{align*}
        \varphi:\quad\quad G\quad &\longrightarrow\quad A\\
        \chi_{_{(a_1,\cdots,a_t)}}&\longmapsto(a_1,\cdots,a_t)
    \end{align*}
It is easy to verify that $\varphi$ is group homomorphism and it's obvious that $\varphi$ is surjective and injective . Hence $G\simeq A$ . 
\end{proof}
\textbf{Corollary 3.3}.(\cite{bib4})The  primitive idempotents of $F_q[G]$ are given by 
\begin{equation}
    v_{_x}=\dfrac{1}{n}\sum_{(a_1,a_2,\cdots,a_t)}(\chi_{_{(a_1,a_2,\cdots,a_t)}}(x))^{\scriptscriptstyle-1}\chi_{_{(a_1,a_2,\cdots,a_t)}}
\end{equation}
one for each $x$ in $A$ .\\
\quad\\
\textbf{Corollary 3.4}. The  primitive idempotents of $F_q[A]$ are given by
\begin{equation}
    e_{_x}=\dfrac{1}{n}\sum_{(a_1,a_2,\cdots,a_t)}(\chi_{_{(a_1,a_2,\cdots,a_t)}}(x))^{\scriptscriptstyle-1}(a_1,a_2,\cdots,a_t)
\end{equation}
one for each $x$ in $A$ .
\begin{proof}
    From \textbf{Corollary 3.2} , $G\simeq A$ . Hence $F_q[G]\simeq F_q[A]$ is $F_q$-algebra isomorphism by performing linear extention of $\varphi$ , the primitive idempotents of $F_q[G]$ and $ F_q[A]$ is one-to-one correspondent by $\varphi$ . By \textbf{Corollary 3.3}, $\{\varphi(v_{_x})|x\in{A}\}$=$\{\varphi(e_{_x})|x\in{A}\}$ are the primitive idempotents of $F_q[A]$ .
\end{proof}
\quad\\
\textbf{Corollary 3.5}. If $A=$\textless $a$\textgreater is cyclic group of order $n$ , let $\xi$ be a primitive $n-th$ root of $F_q$ , then the primitive idempotents of $F_q[A]$ are given by
\begin{align}
    e_{_i}&=\dfrac{1}{n}\sum_{j=0}^{n-1}(\chi_{_{(j)}}(a^{\scriptscriptstyle i}))^{\scriptscriptstyle-1}a^{j}\\
    &=\dfrac{1}{n}\sum_{j=0}^{n-1}\xi^{\scriptscriptstyle-ij}a^{j}\qquad\qquad\qquad\qquad\forall\ 0\leq i\leq n-1
\end{align}
\begin{proof}
    It follows from \textbf{Corollary 3.4} immediately .
\end{proof}
\textbf{Remark.}. It's easy to verify $\sum\limits_{i=0}^{n-1}e_{_i}=1$ , 
$
e_{_i}e_{_j} = \left\{
\begin{array}{ll}
e_{_i} & ,i=j  \\
0 & ,i\neq j
\end{array}
\right.
$\\
\quad\\
\textbf{Corollary 3.6}. If gcd($2n,q$)=1 , $F_q$ contains all the $n-th$ roots of unity , then the central primitive idempotents of $F_qD_{2n}$ are given as follow
\begin{equation}
   \begin{cases}
   (\frac{1+b}{2})e_{_0},(\frac{1-b}{2})e_{_0},(e_{_1}+e_{_{n-1}}),\cdots,(e_{_{\frac{n-1}{2}}}+e_{_{\frac{n+1}{2}}}) & ,\text{$n$ is odd}\\
   (\frac{1+b}{2})e_{_0},(\frac{1-b}{2})e_{_0},(\frac{1+b}{2})e_{_{\frac{n}{2}}},(\frac{1-b}{2})e_{_{\frac{n}{2}}}(e_{_1}+e_{_{n-1}}),\cdots,(e_{_{\frac{n}{2}-1}}+e_{_{\frac{n}{2}+1}}) & ,\text{$n$ is even}
   \end{cases}
\end{equation}
\begin{proof}
    If $n$ is odd , from \textbf{Corollary 2.2} , we know that
    \begin{align*}
        P:F_qD_{2n}&\longrightarrow\qquad\mathop{\bigoplus}\limits_{j=1}^{\delta+\frac{n-\delta}{2}}A_j\\
        a\quad&\longmapsto\begin{pmatrix}
            1&1
        \end{pmatrix}
        \times
        \begin{pmatrix}
            \xi&0\\
            0&\xi^{\scriptscriptstyle-1}
        \end{pmatrix}
        \times\cdots\times
        \begin{pmatrix}
            \xi^{\frac{n-1}{2}}&0\\
            0&\xi^{-\frac{n-1}{2}}
        \end{pmatrix}\\
        b\quad&\longmapsto\begin{pmatrix}
            1&-1
        \end{pmatrix}
        \times
        \begin{pmatrix}
            0&1\\
            1&0
        \end{pmatrix}
         \times\cdots\times
        \begin{pmatrix}
            0&1\\
            1&0
        \end{pmatrix}
    \end{align*}
    It is easy to verify that
    \begin{align*}
    P(e_{_0})=P(\dfrac{1}{n}\sum_{j=1}^{n-1}a^{j})&=\begin{pmatrix}
        1&1
    \end{pmatrix}
    \times
        \begin{pmatrix}
            \dfrac{1}{n}\sum\limits_{j=1}^{n-1}\xi^{\scriptscriptstyle j}&0\\
            0&\dfrac{1}{n}\sum\limits_{j=1}^{n-1}\xi^{\scriptscriptstyle-j}
        \end{pmatrix}
        \times\cdots\times
        \begin{pmatrix}
            \dfrac{1}{n}\sum\limits_{j=1}^{n-1}\xi^{\scriptscriptstyle\frac{n-1}{2}\cdot j}&0\\
            0&\dfrac{1}{n}\sum\limits_{j=1}^{n-1}\xi^{\scriptscriptstyle-\frac{n-1}{2}\cdot j}
        \end{pmatrix}\\
        &=\begin{pmatrix}
          1&1  
        \end{pmatrix}
        \times
        \begin{pmatrix}
            0&0\\
            0&0
        \end{pmatrix}
        \times\cdots\times
        \begin{pmatrix}
            0&0\\
            0&0
        \end{pmatrix}
    \end{align*}
    for $1\leq i\leq\dfrac{n-1}{2}$ , 
    \begin{align*}
        P(e_{_i})&=\begin{pmatrix}
        \dfrac{1}{n}\sum\limits_{j=1}^{n-1}\xi^{\scriptscriptstyle-ij}&\dfrac{1}{n}\sum\limits_{j=1}^{n-1}\xi^{\scriptscriptstyle-ij}
    \end{pmatrix}
    \times
        \begin{pmatrix}
            \dfrac{1}{n}\sum\limits_{j=1}^{n-1}\xi^{\scriptscriptstyle -(i-1)j}&0\\
            0&\dfrac{1}{n}\sum\limits_{j=1}^{n-1}\xi^{\scriptscriptstyle-(i+1)j}
        \end{pmatrix}
        \times\cdots\times
        \begin{pmatrix}
            \dfrac{1}{n}\sum\limits_{j=1}^{n-1}\xi^{-({\scriptscriptstyle i-\frac{n-1}{2})j}}&0\\
            0&\dfrac{1}{n}\sum\limits_{j=1}^{n-1}\xi^{-({\scriptscriptstyle i+\frac{n-1}{2})j}}
        \end{pmatrix}\\
        &=\begin{pmatrix}
          0&0
        \end{pmatrix}
        \times
        \begin{pmatrix}
            0&0\\
            0&0
        \end{pmatrix}
        \times\cdots\times
        \begin{pmatrix}
            1&0\\
            0&0
        \end{pmatrix}
        \times\cdots\times
        \begin{pmatrix}
            0&0\\
            0&0
        \end{pmatrix}
    \end{align*}
    and for $\dfrac{n+1}{2}\leq i\leq n-1$ ,
    \begin{align*}
        P(e_{_i})&=\begin{pmatrix}
        \dfrac{1}{n}\sum\limits_{j=1}^{n-1}\xi^{\scriptscriptstyle-ij}&\dfrac{1}{n}\sum\limits_{j=1}^{n-1}\xi^{\scriptscriptstyle-ij}
    \end{pmatrix}
    \times
        \begin{pmatrix}
            \dfrac{1}{n}\sum\limits_{j=1}^{n-1}\xi^{\scriptscriptstyle -(i-1)j}&0\\
            0&\dfrac{1}{n}\sum\limits_{j=1}^{n-1}\xi^{\scriptscriptstyle-(i+1)j}
        \end{pmatrix}
        \times\cdots\times
        \begin{pmatrix}
            \dfrac{1}{n}\sum\limits_{j=1}^{n-1}\xi^{-({\scriptscriptstyle i-\frac{n-1}{2})j}}&0\\
            0&\dfrac{1}{n}\sum\limits_{j=1}^{n-1}\xi^{-({\scriptscriptstyle i+\frac{n-1}{2})j}}
        \end{pmatrix}\\
        &=\begin{pmatrix}
          0&0
        \end{pmatrix}
        \times
        \begin{pmatrix}
            0&0\\
            0&0
        \end{pmatrix}
        \times\cdots\times
        \begin{pmatrix}
            0&0\\
            0&1
        \end{pmatrix}
        \times\cdots\times
        \begin{pmatrix}
            0&0\\
            0&0
        \end{pmatrix}
    \end{align*}
    where for the case $1\leq i\leq\dfrac{n-1}{2}$ , $1$ only appears in position $(1,1)$ of the $(i+1)-th$ factor $\mathop{\bigoplus}\limits_{j=1}^{\delta+\frac{n-\delta}{2}}A_j$ and $0$ appears in the rest . For the case $\dfrac{n+1}{2}\leq i\leq n-1$ , $1$ only appears in position $(2,2)$ of the $(n-i+1)-th$ factor $\mathop{\bigoplus}\limits_{j=1}^{\delta+\frac{n-\delta}{2}}A_j$ and $0$ appears in the rest . If $n$ is even , the discussion is similar .
    Hence , the central primitive idempotents of $F_qD_{2n}$ are given by (13) .
\end{proof}
\textbf{Remark}. Let $R=F_qD_{2n}$ . Under the condition of this section , the irreducible representation of $D_{2n}$ is given by
\[
\text{$n$ is odd:} \begin{cases}
  \text{1-dimension:}\{R(\frac{1\pm b}{2}e_{_0})\} \\
  \text{2-dimension:}\{Re_{_i}\}\quad,\quad1\leq i\leq n-1
\end{cases} \:
\text{$n$ is even:} \begin{cases}
  \text{1-dimension:}\{R(\frac{1\pm b}{2}e_{_0})\;\text{or}\;R(\frac{1\pm b}{2}e_{_{\frac{n}{2}}})\} \\
  \text{2-dimension:}\{Re_{_i}\}\quad,\quad1\leq i\leq n-1,n\neq\dfrac{n}{2}
\end{cases}
\]
up to isomorphism . Moreover , $Re_{_i}\simeq Re_{_{n-i}}$ by $f$ as follow ,
    \begin{align*}
        f:Re_{_i}&\longrightarrow Re_{_{n-i}}\\
        e_{_i}&\longmapsto be_{_{n-i}}\\
        be_{_i}&\longmapsto e_{_{n-i}}
    \end{align*}
for $1\leq i\leq\dfrac{n-1}{2}$ if $n$ is odd  and $1\leq i\leq\dfrac{n}{2}-1$ if $n$ is even .
\section{Some MDS codes over dihedral group code}
In this section , let $D_{2n}$ be a dihedral group where $n$ is odd . $F_q$ denotes a finite group such that $gcd(2n,q)=1$ , $2n|q-1$ and $2n<q-1$ . Let $\eta$ denotes $2n-th$ primitive root of unity . From (5)(6) , $F_qD_{2n}\simeq\mathop{\bigoplus}\limits_{j=1}^{1+\frac{n-1}{2}}A_j$ . For full matrix algebra of order $2$ over $F_q$ , the determinant of any proper left ideal $I$ is $0$ .Then the form of $I$ is given as follow(\cite{bib2})
\begin{equation}
    I=\left\{\begin{pmatrix}
           kx&ky\\
           tx&ty
      \end{pmatrix}{\Bigg|}
      \quad k,t\in{F_q} \right\}
      =M_2(F_q)\begin{pmatrix}
    x&y\\
    0&0
\end{pmatrix}
\end{equation}
where $(x,y)\neq(0,0)$ . Let
\[
I_{_j}(x,y):=\begin{cases}
  \quad {A_{_1}(x,y)}\quad&,j=1\\
  \left\{\begin{pmatrix}
           kx&ky\\
           tx&ty
      \end{pmatrix}{\Bigg|}
      \quad k,t\in{F_q} \right\}
       =A_{_j}\begin{pmatrix}
    x&y\\
    0&0
\end{pmatrix}
      \quad&,1\leq j\leq\dfrac{n-1}{2}
\end{cases}
\]
\\
\textbf{Lemma 4.1}.(\cite{bib2},\textbf{Theorem 3}) Consider the decomposition of (6) . Then for any code $I\subset R$ we have
\[
P(I)=\bigoplus\limits_{j=1}^{1+\frac{n-1}{2}}B_{_j},\qquad 
    B_{_j}=\begin{cases}
    A_{_j}&,j\in{J_{_1}}\\
    I_{_j}(0,1)&,j\in{J_{_2}}\\
    I_{_j}(1,0)&,j\in{J_{_3}}\\
    I_{_j}(1,x_{_j})&,j\in{J_{_4}}\\
    0&,j\notin {J_{_1}}\cup{J_{_2}}\cup{J_{_3}}\cup{J_{_4}}
\end{cases}
\]
On the other hand , for any $B_{_j}$ such that $B_{_j}=A_{_j}$ or $B_{_j}=I_{_j}(x_{_j},y_{_j})$ or $B_{_j}=0$ , $P^{\scriptscriptstyle-1}(\bigoplus\limits_{j=1}^{1+\frac{n-1}{2}}B_{_j})$ is a $D_{_{2n}}$-code , where $x_{_j},y_{_j}\in{F_q}$ .
\begin{proof}
    Since $F_qD_{_{2n}}$ and $\mathop{\bigoplus}\limits_{j=1}^{1+\frac{n-1}{2}}A_j$ are isomorphic by $P$ , it follows that there is one-to-one correspondence between the $D_{_{2n}}$-codes and the left ideals of $\mathop{\bigoplus}\limits_{j=1}^{1+\frac{n-1}{2}}A_j$ . As is well-known , any left ideal of a direct sum of algebras with identity is a direct sum if ideals of the direct summands . And it's easy to verify any ideal of $A_{_j}$ is of the form $B_{_j}$ as mentioned above . Hence , the lemma is proved . 
\end{proof}
From the proof of \textbf{Corollary 3.6} , we know that $P(Re_{_0})=A_{_1}$ . And for $1\leq j\leq\dfrac{n-1}{2}$ , $P(Re_{_j}\oplus Re_{_{n-j}})=A_{_{j+1}}$\\
\quad\\
\textbf{Theorem 4.2}. Let $I=\mathop{\bigoplus}\limits_{j\neq 1,n-1}Re_{_j}\oplus R(e_{_1}+\eta be_{_{n-1}})$ is a left ideal of $R$ , then $dim_{_{F_q}}I=2n-2$ , $\phi(I)$ is $[2n,2n-2,3]$ MDS code .
\begin{proof}
    From (13) , we know that $e_{_i}=\dfrac{1}{n}\sum\limits_{j=0}^{n-1}\xi^{\scriptscriptstyle-ij}a^{j}$ . For any $0\leq t\leq n-1$
    \begin{align*}
        a^{\scriptscriptstyle t}e_{_i}&=a^{\scriptscriptstyle t}(\dfrac{1}{n}\sum\limits_{j=1}^{n-1}\xi^{\scriptscriptstyle-ij}a^{j})=
        \dfrac{1}{n}\sum\limits_{j=1}^{n-1}\xi^{\scriptscriptstyle-ij}a^{\scriptscriptstyle(j+t)}=\xi^{\scriptscriptstyle it}(\dfrac{1}{n}\sum\limits_{j=1}^{n-1}\xi^{\scriptscriptstyle-i(j+t)}a^{\scriptscriptstyle(j+t)})=\xi^{\scriptscriptstyle it}e_{_i}\\
        ba^{\scriptscriptstyle t}e_{_i}&=a^{\scriptscriptstyle t}(\dfrac{1}{n}\sum\limits_{j=1}^{n-1}\xi^{\scriptscriptstyle-ij}a^{j})=\xi^{\scriptscriptstyle it}be_{_i}
    \end{align*}
Therefore , \{$e_{_i},be_{_i}$\} is a set of basis of $Re_{_i}$ and $dim_{_{F_q}}Re_{_i}=2$ . Likewise , 
\begin{align*}
    a^{\scriptscriptstyle t}(e_{_1}+\eta be_{_{n-1}})&=a^{\scriptscriptstyle t}e_{_1}+\eta ba^{\scriptscriptstyle n-t}e_{_{n-1}}=\xi^{\scriptscriptstyle t}e_{_1}+\eta\xi^{\scriptscriptstyle (n-1)(n-t)}be_{_{n-1}}=\xi^{\scriptscriptstyle t}(e_{_1}+\eta be_{_{n-1}})\\
    ba^{\scriptscriptstyle t}(e_{_1}+\eta be_{_{n-1}})&=ba^{\scriptscriptstyle t}e_{_1}+\eta ba^{\scriptscriptstyle t}be_{_{n-1}}=ba^{\scriptscriptstyle t}e_{_1}+\eta a^{\scriptscriptstyle n-t}e_{_{n-1}}=\xi^{\scriptscriptstyle t}(be_{_1}+\eta e_{_{n-1}})
\end{align*}
Hence , $\{e_{_1}+\eta be_{_{n-1}},be_{_1}+\eta e_{_{n-1}}\}$ is a set of basis of $R(e_{_1}+\eta e_{_{n-1}})$ and $dim_{_{F_q}}R(e_{_1}+\eta e_{_{n-1}})=2$. By the calculation above , it's obvious $dim_{_{F_q}}I=2n-2$ . Hence , $\phi(I)$ is $[2n,2n-2,d]$ group code . It is well-known that the parameters $[n,k,d]$ of any linear code over $F_q$ satisfy $d\leq n-k+1$ i.e. \textbf{Singleton bound} , so we only need to verify the distance of $\phi(I)$ is not equal to $1$ or $2$ .  Use the notation above in \textbf{Lemma 4.1} , from the proof of \textbf{Corollary 3.6} , we derive
\[
P(I)=\bigoplus\limits_{j=1}^{1+\frac{n-1}{2}}B_{_j},\qquad 
    B_{_j}=\begin{cases}
    A_{_j}&,j\in{\{1,3,4\cdots,1+\dfrac{n-1}{2}\}}\\
    I_{_j}(1,\eta)&,j=2
\end{cases}
\]
If $d(\phi(I))=1$ , there exist $u\in{I}$ such that $supp(u)=1$ . That is $u=\lambda_i a^{\scriptscriptstyle i}\text{ or }\lambda_i ba^{\scriptscriptstyle i}$ , where $0\leq i\leq n-1$ and $\lambda_i\neq0$ . First , we consider $u=\lambda_i a^{\scriptscriptstyle i}$ , from \textbf{Corollary 2.2} ,
\begin{equation*}
    P(u)=\begin{pmatrix}
          \lambda_i&\lambda_i
        \end{pmatrix}
        \times
        \begin{pmatrix}
            \lambda_i\xi^{\scriptscriptstyle i}&0\\
            0&\lambda_i\xi^{\scriptscriptstyle-i}
        \end{pmatrix}
        \times\cdots\times
        \begin{pmatrix}
            \lambda_i\xi^{(\scriptscriptstyle\frac{n-1}{2})i}&0\\
            0&\lambda_i\xi^{(\scriptscriptstyle-\frac{n-1}{2})i}
        \end{pmatrix}
\end{equation*}
where $\begin{pmatrix}
            \lambda_i\xi^{\scriptscriptstyle i}&0\\
            0&\lambda_i\xi^{\scriptscriptstyle-i}
        \end{pmatrix}\in{I_{_2}}(1,\eta)$ . 
However , the determinant of any elements of $I_{_2}(1,\eta)$ is $0$ and 
$\begin{vmatrix}
\lambda_i\xi^{\scriptscriptstyle i}&0\\
0&\lambda_i\xi^{\scriptscriptstyle-i} \\
\end{vmatrix}=\lambda_{i}^{\scriptscriptstyle2}\neq0$ . 
It's a contradiction . The discussion is the same when $u=\lambda_i ba^{\scriptscriptstyle i}$.\\
\quad\\
If $d(\phi(I))=2$ , there exist $u\in{I}$ such that $supp(u)=2$ . That is $u$ is divided into three cases:
\[
u=\begin{cases}
\lambda_{i}a^{\scriptscriptstyle i}+\lambda_{j}a^{\scriptscriptstyle j}&,0\leq i<j\leq n-1 \text{ and } \lambda_i,\lambda_j\neq0\\
\lambda_{i}a^{\scriptscriptstyle i}+\lambda_{j}ba^{\scriptscriptstyle j}&,\lambda_i,\lambda_j\neq0\\
\lambda_{i}ba^{\scriptscriptstyle i}+\lambda_{j}ba^{\scriptscriptstyle j}&,0\leq i<j\leq n-1\text{ and } \lambda_i,\lambda_j\neq0
\end{cases}
\]
case 1: If $u=\lambda_{i}a^{\scriptscriptstyle i}+\lambda_{j}a^{\scriptscriptstyle j}\:,\:0\leq i<j\leq n-1 \text{ and } \lambda_i,\lambda_j\neq0$ , we have 
\begin{equation*}
    P(u)=\begin{pmatrix}
          \lambda_i+\lambda_j&\lambda_i+\lambda_j
        \end{pmatrix}
        \times
        \begin{pmatrix}
            \lambda_i\xi^{\scriptscriptstyle i}+\lambda_j\xi^{\scriptscriptstyle j}&0\\
            0&\lambda_i\xi^{\scriptscriptstyle-i}+\lambda_j\xi^{\scriptscriptstyle-j}
        \end{pmatrix}
        \times\cdots\times
        \begin{pmatrix}
            \lambda_i\xi^{\scriptscriptstyle(\frac{n-1}{2})i}+\lambda_j\xi^{\scriptscriptstyle(\frac{n-1}{2})j}&0\\
            0&\lambda_i\xi^{(\scriptscriptstyle-\frac{n-1}{2})i}+\lambda_j\xi^{(\scriptscriptstyle-\frac{n-1}{2})j}
        \end{pmatrix}
\end{equation*}
where $\begin{pmatrix}
           \lambda_i\xi^{\scriptscriptstyle i}+\lambda_j\xi^{\scriptscriptstyle j}&0\\
            0&\lambda_i\xi^{\scriptscriptstyle-i}+\lambda_j\xi^{\scriptscriptstyle-j}
        \end{pmatrix}\in{I_{_2}}(1,\eta)$ . 
Therefore , 
$
\left\{
\begin{array}{ll}
\lambda_i\xi^{\scriptscriptstyle i}+\lambda_j\xi^{\scriptscriptstyle j}&=0\\
\lambda_i\xi^{\scriptscriptstyle-i}+\lambda_j\xi^{\scriptscriptstyle-j}&=0
\end{array}
\right.
$\\
\quad\\
In other words , there exist $\lambda_i,\lambda_j\neq0$ such that the simultaneous equations mentioned above established . Then we can derive 
$\begin{vmatrix}
\xi^{\scriptscriptstyle i}&\xi^{\scriptscriptstyle j}\\
\xi^{\scriptscriptstyle -i}&\xi^{\scriptscriptstyle -j}
\end{vmatrix}=\xi^{\scriptscriptstyle i-j}-\xi^{\scriptscriptstyle j-i}=0\Longrightarrow\xi^{\scriptscriptstyle i-j}=\xi^{\scriptscriptstyle j-i}\Longrightarrow\xi^{\scriptscriptstyle 2(i-j)}=1$ .
Therefore , we derive $n|2(i-j)$ . By $n$ is odd , then $n|(i-j)$ . But from $0\leq i<j\leq n-1$ , that's impossible . \\
case 2: If $\lambda_{i}a^{\scriptscriptstyle i}+\lambda_{j}ba^{\scriptscriptstyle j}\:,\:\lambda_i,\lambda_j\neq0$ , we have 
\begin{equation*}
    P(u)=\begin{pmatrix}
          \lambda_i+\lambda_j&\lambda_i-\lambda_j
        \end{pmatrix}
        \times
        \begin{pmatrix}
            \lambda_i\xi^{\scriptscriptstyle i}&\lambda_j\xi^{\scriptscriptstyle-j}\\
            \lambda_j\xi^{\scriptscriptstyle j}&\lambda_i\xi^{\scriptscriptstyle-i}
        \end{pmatrix}
        \times\cdots\times
        \begin{pmatrix}
            \lambda_i\xi^{\scriptscriptstyle(\frac{n-1}{2})i}&\lambda_j\xi^{(\scriptscriptstyle-\frac{n-1}{2})j}\\
            \lambda_j\xi^{\scriptscriptstyle(\frac{n-1}{2})j}&\lambda_i\xi^{(\scriptscriptstyle-\frac{n-1}{2})i}
        \end{pmatrix}
\end{equation*}
where $\begin{pmatrix}
            \lambda_i\xi^{\scriptscriptstyle i}&\lambda_j\xi^{\scriptscriptstyle-j}\\
            \lambda_j\xi^{\scriptscriptstyle j}&\lambda_i\xi^{\scriptscriptstyle-i}
        \end{pmatrix}\in{I_{_2}}(1,\eta)$ . 
Therefore , 
$\begin{vmatrix}
            \lambda_i\xi^{\scriptscriptstyle i}&\lambda_j\xi^{\scriptscriptstyle-j}\\
            \lambda_j\xi^{\scriptscriptstyle j}&\lambda_i\xi^{\scriptscriptstyle-i}
\end{vmatrix}=(\lambda_i-\lambda_j)(\lambda_i+\lambda_j)=0$ .\\
\quad\\
We derive $\lambda_i=\lambda_j$ or $\lambda_i=-\lambda_j$ . For $\lambda_i=\lambda_j$ , 
\begin{equation*}
    \begin{pmatrix}
        \lambda_i\xi^{\scriptscriptstyle i}&\lambda_j\xi^{\scriptscriptstyle-j}\\
        \lambda_j\xi^{\scriptscriptstyle j}&\lambda_i\xi^{\scriptscriptstyle-i}
    \end{pmatrix}
    =
    \begin{pmatrix}
        \lambda_i\xi^{\scriptscriptstyle i}&\lambda_i\xi^{\scriptscriptstyle-j}\\
        \lambda_i\xi^{\scriptscriptstyle j}&\lambda_i\xi^{\scriptscriptstyle-i}
    \end{pmatrix}
    =\lambda_i\xi^{\scriptscriptstyle i}
    \begin{pmatrix}
        1&\xi^{\scriptscriptstyle-(i+j)}\\
        0&0
    \end{pmatrix}
    +\lambda_i\xi^{\scriptscriptstyle j}
    \begin{pmatrix}
        0&0\\
        1&\xi^{-\scriptscriptstyle(i+j)}
    \end{pmatrix}\in{I_{_2}}(1,\eta)
\end{equation*}
However , $\xi^{-\scriptscriptstyle(i+j)}$ is $n-th$ root of unity , from $2n<q-1$ and $\eta$ is $ (q-1)-th$ primitive root of unity we deduce $\xi^{-\scriptscriptstyle(i+j)}\neq\eta$ . Hence , for any $0\leq i,j\leq n-1$ , 
\begin{equation*}
    \begin{pmatrix}
        \lambda_i\xi^{\scriptscriptstyle i}&\lambda_j\xi^{\scriptscriptstyle-j}\\
        \lambda_j\xi^{\scriptscriptstyle j}&\lambda_i\xi^{\scriptscriptstyle-i}
    \end{pmatrix}\notin{I_{_2}}(1,\eta)
\end{equation*}
For $\lambda_i=-\lambda_j$ , it is similar . That is $-\xi^{-\scriptscriptstyle(i+j)}$ is $2n-th$ root of unity , from $2n<q-1$ and $\eta$ is $ (q-1)-th$ primitive root of unity we deduce $-\xi^{-\scriptscriptstyle(i+j)}\neq\eta$. It's a contradiction .\\
case 3: If $u=\lambda_{i}ba^{\scriptscriptstyle i}+\lambda_{j}ba^{\scriptscriptstyle j}\:,\:0\leq i<j\leq n-1 \text{ and } \lambda_i,\lambda_j\neq0$ . The discussion is the same as case 1 by the $F_q-$algebra isomorphism given in \textbf{Corollary 2.2} .\\
In conclusion , $\phi(I)$ is $[2n,2n-2,3]$ group code .
\end{proof}
\textbf{Corollary 4.3}. Let $I=\mathop{\bigoplus}\limits_{j\neq 1,n-1}Re_{_j}\oplus R(e_{_1}+\eta be_{_{n-1}})$ , The generate matrix of $\phi(I)$ is 
\begin{equation*}
    G_{I}=\begin{pmatrix}
        \phi(e_{_0})\\
        \phi(be_{_0})\\
        \phi(e_{_1}+\eta be_{_{n-1}})\\
        \phi(be_{_1}+\eta e_{_{n-1}})\\
        \phi(e_{_2})\\
        \phi(be_{_2})\\
        \vdots\\
        \phi(e_{_{n-2}})\\
        \phi(be_{_{n-2}})
    \end{pmatrix}
    =\begin{pmatrix}
        1&1&\cdots&1&0&0&\cdots&0\\
        0&0&\cdots&0&1&1&\cdots&1\\
        1&\xi^{\scriptscriptstyle-1}&\cdots&\xi^{\scriptscriptstyle-(n-1)}&\eta&\eta\xi^{\scriptscriptstyle1}&\cdots&\eta\xi^{\scriptscriptstyle n-1}\\
        \eta&\eta\xi^{\scriptscriptstyle1}&\cdots&\eta\xi^{\scriptscriptstyle n-1}&1&\xi^{\scriptscriptstyle-1}&\cdots&\xi^{\scriptscriptstyle-(n-1)}\\
        1&\xi^{\scriptscriptstyle-2}&\cdots&\xi^{\scriptscriptstyle-2(n-1)}&0&0&\cdots&0\\
        0&0&\cdots&0&1&\xi^{\scriptscriptstyle-2}&\cdots&\xi^{\scriptscriptstyle-2(n-1)}\\
        \vdots&\vdots&\quad&\vdots&\vdots&\vdots&\quad&\vdots\\
        1&\xi^{\scriptscriptstyle-(n-2))}&\cdots&\xi^{\scriptscriptstyle-(n-2)(n-1)}&0&0&\cdots&0\\
        0&0&\cdots&0&1&\xi^{\scriptscriptstyle-(n-2)}&\cdots&\xi^{\scriptscriptstyle-(n-2)(n-1)}
    \end{pmatrix}
\end{equation*}
\textbf{Corollary 4.4}. If $\beta\in{F_q^{*}}$ such that $ord(\beta)>2n$ , then $\phi(I)$ is $[2n,2n-2,3]$ MDS dihedral group code , where $I=\mathop{\bigoplus}\limits_{j\neq 1,n-1}Re_{_j}\oplus R(e_{_1}+\beta be_{_{n-1}})$ .
\begin{proof}
    It follows from the proof of \textbf{Theorem 4.2} immediately .
\end{proof}
\textbf{Corollary 4.5}. If $\beta\in{F_q^{*}}$ such that $ord(\beta)>2n$ and for $1\leq s\leq\dfrac{n-1}{2}$ such that $gcd(s,n)=1$ , then
$\phi(I)$ is $[2n,2n-2,3]$ MDS dihedral group code , where $I=\mathop{\bigoplus}\limits_{j\neq i,n-i}Re_{_j}\oplus R(e_{_s}+\beta be_{_{n-s}})$
\begin{proof}
From the proof of \textbf{Theorem 4.2} ,
\[
P(I)=\bigoplus\limits_{j=1}^{1+\frac{n-1}{2}}B_{_j},\qquad 
    B_{_j}=\begin{cases}
    A_{_j}&,j\in{\{1,\cdots,s,s+2,\cdots,1+\dfrac{n-1}{2}\}}\\
    I_{_j}(1,\eta)&,j=s+1
\end{cases}
\]
    case 1: If $u=\lambda_{i}a^{\scriptscriptstyle i}+\lambda_{j}a^{\scriptscriptstyle j}\:,\:0\leq i<j\leq n-1 \text{ and } \lambda_i,\lambda_j\neq0$ , the component of $P(u)$ in $P(I)$ is \\
    \quad\\
    given by 
       $\begin{pmatrix}
           \lambda_i\xi^{\scriptscriptstyle is}+\lambda_j\xi^{\scriptscriptstyle js}&0\\
            0&\lambda_i\xi^{\scriptscriptstyle-is}+\lambda_j\xi^{\scriptscriptstyle-js}
        \end{pmatrix}\in{I_{_{s+1}}}(1,\beta)$ . 
   Therefore , 
$
\left\{
\begin{array}{ll}
\lambda_i\xi^{\scriptscriptstyle is}+\lambda_j\xi^{\scriptscriptstyle js}&=0\\
\lambda_i\xi^{\scriptscriptstyle-is}+\lambda_j\xi^{\scriptscriptstyle-js}&=0
\end{array}
\right.
$\\
\quad\\
 Then we can derive 
$\begin{vmatrix}
\xi^{\scriptscriptstyle is}&\xi^{\scriptscriptstyle js}\\
\xi^{\scriptscriptstyle-is}&\xi^{\scriptscriptstyle-js}
\end{vmatrix}=\xi^{\scriptscriptstyle(i-j)s}-\xi^{-\scriptscriptstyle(i-j)s}=0\Longrightarrow\xi^{\scriptscriptstyle(i-j)s}=\xi^{\scriptscriptstyle(j-i)s}\Longrightarrow\xi^{\scriptscriptstyle 2(i-j)s}=1$\\
\quad\\
Therefore , we derive $n|2(i-j)s$ . By $n$ is odd and $(s,n)=1$ , then $n|(i-j)$ . But from $0\leq i<j\leq n-1$ , that's impossible . \\ 
case 2: If $\lambda_{i}a^{\scriptscriptstyle i}+\lambda_{j}ba^{\scriptscriptstyle j}\:,\:\lambda_i,\lambda_j\neq0$ , the component of $P(u)$ in $P(I)$ is 
$\begin{pmatrix}
    \lambda_i\xi^{\scriptscriptstyle is}&\lambda_j\xi^{\scriptscriptstyle-js}\\
    \lambda_j\xi^{\scriptscriptstyle js}&\lambda_i\xi^{\scriptscriptstyle-is}
\end{pmatrix}$ .\\
\quad\\
Then we can derive 
$\begin{vmatrix}
    \lambda_i\xi^{\scriptscriptstyle is}&\lambda_j\xi^{\scriptscriptstyle-js}\\
    \lambda_j\xi^{\scriptscriptstyle js}&\lambda_i\xi^{\scriptscriptstyle-is}
\end{vmatrix}=(\lambda_i-\lambda_j)(\lambda_i+\lambda_j)=0$ . For $\lambda_i=\lambda_j$ , 
\begin{equation*}
    \begin{pmatrix}
        \lambda_i\xi^{\scriptscriptstyle is}&\lambda_j\xi^{\scriptscriptstyle-js}\\
        \lambda_j\xi^{\scriptscriptstyle js}&\lambda_i\xi^{\scriptscriptstyle-is}
    \end{pmatrix}
    =
    \begin{pmatrix}
        \lambda_i\xi^{\scriptscriptstyle is}&\lambda_i\xi^{\scriptscriptstyle-js}\\
        \lambda_i\xi^{\scriptscriptstyle js}&\lambda_i\xi^{\scriptscriptstyle-is}
    \end{pmatrix}
    =\lambda_i\xi^{\scriptscriptstyle is}
    \begin{pmatrix}
        1&\xi^{\scriptscriptstyle-(i+j)s}\\
        0&0
    \end{pmatrix}
    +\lambda_i\xi^{\scriptscriptstyle js}
    \begin{pmatrix}
        0&0\\
        1&\xi^{-\scriptscriptstyle(i+j)s}
    \end{pmatrix}\in{I_{_{s+1}}}(1,\beta)
\end{equation*}
However , $\xi^{-\scriptscriptstyle(i+j)s}$ is $n-th$ root of unity , from $ord(\beta)>2n$ we deduce $\xi^{-\scriptscriptstyle(i+j)s}\neq\beta$ . Hence , for any $0\leq i,j\leq n-1$ , 
\begin{equation*}
    \begin{pmatrix}
        \lambda_i\xi^{\scriptscriptstyle is}&\lambda_j\xi^{\scriptscriptstyle-js}\\
        \lambda_j\xi^{\scriptscriptstyle js}&\lambda_i\xi^{\scriptscriptstyle-is}
    \end{pmatrix}\notin{I_{_{s+1}}}(1,\beta)
\end{equation*}
For $\lambda_i=-\lambda_j$ , it is similar .\\
case 3: If $u=\lambda_{i}ba^{\scriptscriptstyle i}+\lambda_{j}ba^{\scriptscriptstyle j}\:,\:0\leq i<j\leq n-1 \text{ and } \lambda_i,\lambda_j\neq0$ . The discussion is the same as case 1 .
\end{proof}
\textbf{Theorem 4.6}. Let $I=\mathop{\bigoplus}\limits_{j\neq 0,1,n-1}Re_{_j}\oplus R(\dfrac{1-b}{2}e_{_0})\oplus R(e_{_1}+\eta be_{_{n-1}})$ is a left ideal of $R$ , then $dim_{_{F_q}}I=2n-3$ , $\phi(I)$ is $[2n,2n-3,4]$ MDS code .
\begin{proof}
From \textbf{Corollary 3.6} and \textbf{Theorem 4.2} , we know that $dim_{_{F_q}}I=2n-3$ . Then $\phi(I)$ is $[2n,2n-3,d]$ group code . By \textbf{Singleton bound} , we derive $d\leq 4$ . In order to prove $d=4$ , we need to verify that $d\neq 1,2,3$ . The situation that $d\neq 1,2$ is proved in \textbf{Theorem 4.2} , then we only need to verify $d\neq3$ . Use the notation above in \textbf{Lemma 4.1} , we derive
\[
P(I)=\bigoplus\limits_{j=1}^{1+\frac{n-1}{2}}B_{_j},\qquad 
    B_{_j}=\begin{cases}
    A_{_j}&,j\in{\{3,4\cdots,1+\dfrac{n-1}{2}\}}\\
    I_{_j}(0,1)&,j=1\\
    I_{_j}(1,\eta)&,j=2
\end{cases}
\]
If $d(\phi(I))=3$ , there exist $u\in{I}$ such that $supp(u)=3$ . That is $u$ is divided into four cases:
\[
u=\begin{cases}
\lambda_{i}a^{\scriptscriptstyle i}+\lambda_{j}a^{\scriptscriptstyle j}+\lambda_{k}a^{\scriptscriptstyle k}&,0\leq i<j<k\leq n-1 \text{ and } \lambda_i,\lambda_j,\lambda_k\neq0\\
\lambda_{i}a^{\scriptscriptstyle i}+\lambda_{i}a^{\scriptscriptstyle i}+\lambda_{k}ba^{\scriptscriptstyle k}&,0\leq i<j\leq n-1\text{ and }\lambda_i,\lambda_j,\lambda_k\neq0\\
\lambda_{i}a^{\scriptscriptstyle i}+\lambda_{j}ba^{\scriptscriptstyle j}+\lambda_{k}ba^{\scriptscriptstyle k}&,0\leq j<k\leq n-1\text{ and } \lambda_i,\lambda_j,\lambda_k\neq0\\
\lambda_{i}ba^{\scriptscriptstyle i}+\lambda_{j}ba^{\scriptscriptstyle j}+\lambda_{j}ba^{\scriptscriptstyle k}&,0\leq i<j<k\leq n-1 \text{ and } \lambda_i,\lambda_j,\lambda_k\neq0\\
\end{cases}
\]
case 1: If $u=\lambda_{i}a^{\scriptscriptstyle i}+\lambda_{j}a^{\scriptscriptstyle j}+\lambda_{k}a^{\scriptscriptstyle k}\:,\:0\leq i<j<k\leq n-1 \text{ and } \lambda_i,\lambda_j,\lambda_k\neq0$
\begin{align*}
    P(u)=\begin{pmatrix}
          \lambda_i+\lambda_j+\lambda_k\quad,&\lambda_i+\lambda_j+\lambda_k
        \end{pmatrix}
        \times
        \begin{pmatrix}
            \lambda_i\xi^{\scriptscriptstyle i}+\lambda_j\xi^{\scriptscriptstyle j}+\lambda_k\xi^{\scriptscriptstyle k}&0\\
            0&\lambda_i\xi^{\scriptscriptstyle-i}+\lambda_j\xi^{\scriptscriptstyle-j}+\lambda_k\xi^{\scriptscriptstyle-k}
        \end{pmatrix}
        \times\cdots\times
\end{align*} 
Then ,
\begin{align*}
        \begin{pmatrix}
           \lambda_i+\lambda_j+\lambda_k&\lambda_i+\lambda_j+\lambda_k 
        \end{pmatrix}\in{I_{_1}(0,1)}
        \:,\:
        \begin{pmatrix}
             \lambda_i\xi^{\scriptscriptstyle i}+\lambda_j\xi^{\scriptscriptstyle j}+\lambda_k\xi^{\scriptscriptstyle k}&0\\
            0&\lambda_i\xi^{\scriptscriptstyle-i}+\lambda_j\xi^{\scriptscriptstyle-j}+\lambda_k\xi^{\scriptscriptstyle-k}
        \end{pmatrix}\in{I_{_2}(1,\eta)}
\end{align*}
Therefore , 
$
\left\{
\begin{array}{lll}
\lambda_i+\lambda_j+\lambda_k&=0\\
\lambda_i\xi^{\scriptscriptstyle i}+\lambda_j\xi^{\scriptscriptstyle j}+\lambda_k\xi^{\scriptscriptstyle k}&=0\\
\lambda_i\xi^{\scriptscriptstyle-i}+\lambda_j\xi^{\scriptscriptstyle-j}+\lambda_k\xi^{\scriptscriptstyle-k}&=0
\end{array}
\right.
\Longrightarrow\begin{vmatrix}
    1&1&1\\
    \xi^{\scriptscriptstyle i}&\xi^{\scriptscriptstyle j}&\xi^{\scriptscriptstyle k}\\
    \xi^{\scriptscriptstyle-i}&\xi^{\scriptscriptstyle-j}&\xi^{\scriptscriptstyle-k}
\end{vmatrix}=0$ . But ,
\begin{align*}
    \begin{vmatrix}
    1&1&1\\
    \xi^{\scriptscriptstyle i}&\xi^{\scriptscriptstyle j}&\xi^{\scriptscriptstyle k}\\
    \xi^{\scriptscriptstyle-i}&\xi^{\scriptscriptstyle-j}&\xi^{\scriptscriptstyle-k}
\end{vmatrix}=
\begin{vmatrix}
    1&1&1\\
    1&\xi^{\scriptscriptstyle (j-i)}&\xi^{\scriptscriptstyle (k-i)}\\
    1&\xi^{\scriptscriptstyle-(j-i)}&\xi^{\scriptscriptstyle-(k-i)}
\end{vmatrix}&=
\begin{vmatrix}
    1&1&1\\
    0&\xi^{\scriptscriptstyle (j-i)}-1&\xi^{\scriptscriptstyle (k-i)}-1\\
    0&\xi^{\scriptscriptstyle-(j-i)}-1&\xi^{\scriptscriptstyle-(k-i)}-1
\end{vmatrix}\\
&=(\xi^{\scriptscriptstyle (j-i)}-1)(\xi^{\scriptscriptstyle-(k-i)}-1)-(\xi^{\scriptscriptstyle (k-i)}-1)(\xi^{\scriptscriptstyle-(j-i)}-1)\\
&=0
\end{align*}
Then $(\xi^{\scriptscriptstyle (j-i)}-1)(\xi^{\scriptscriptstyle-(k-i)}-1)=(\xi^{\scriptscriptstyle (k-i)}-1)(\xi^{\scriptscriptstyle-(j-i)}-1)$ . From $0\leq i<j<k\leq n-1$ , the factor of the equation is non-zero.
\begin{align*}
    (\xi^{\scriptscriptstyle (j-i)}-1)(\xi^{\scriptscriptstyle-(k-i)}-1)=(\xi^{\scriptscriptstyle (k-i)}-1)(\xi^{\scriptscriptstyle-(j-i)}-1)&\Rightarrow(\xi^{\scriptscriptstyle (j-i)}-1)(\xi^{\scriptscriptstyle-(k-i)}-1)=\xi^{\scriptscriptstyle (k-i)}\xi^{\scriptscriptstyle-(j-i)}(1-\xi^{\scriptscriptstyle -(k-i)})(1-\xi^{\scriptscriptstyle(j-i)})\\
    &\Rightarrow(\xi^{\scriptscriptstyle (j-i)}-1)(\xi^{\scriptscriptstyle-(k-i)}-1)=\xi^{\scriptscriptstyle (k-j)}(\xi^{\scriptscriptstyle -(k-i)}-1)(\xi^{\scriptscriptstyle(j-i)}-1)\\
    &\Rightarrow1=\xi^{\scriptscriptstyle(k-j)}
\end{align*}
It is contradictory to $0\leq i<j<k\leq n-1$ and $\xi$ is $n-th$ primitive of unity .\\
\quad\\
case 2: If $\lambda_{i}a^{\scriptscriptstyle i}+\lambda_{i}a^{\scriptscriptstyle i}+\lambda_{k}ba^{\scriptscriptstyle k}\:,\:0\leq i<j\leq n-1\text{ and }\lambda_i,\lambda_j,\lambda_k\neq0$ . 
\begin{align*}
    P(u)=\begin{pmatrix}
          \lambda_i+\lambda_j+\lambda_k\quad,&\lambda_i+\lambda_j-\lambda_k
        \end{pmatrix}
        \times
        \begin{pmatrix}
             \lambda_i\xi^{\scriptscriptstyle i}+\lambda_j\xi^{\scriptscriptstyle j}&\lambda_k\xi^{\scriptscriptstyle-k}\\
            \lambda_k\xi^{\scriptscriptstyle k}&\lambda_i\xi^{\scriptscriptstyle-i}+\lambda_j\xi^{\scriptscriptstyle-j}
        \end{pmatrix}
        \times\cdots\times\\
\end{align*}
Then ,\\
    $\begin{pmatrix}
           \lambda_i+\lambda_j+\lambda_k\quad,&\lambda_i+\lambda_j-\lambda_k 
        \end{pmatrix}\in{I_{_1}(0,1)}
        \:,\:
        \begin{pmatrix}
             \lambda_i\xi^{\scriptscriptstyle i}+\lambda_j\xi^{\scriptscriptstyle j}&\lambda_k\xi^{\scriptscriptstyle-k}\\
            \lambda_k\xi^{\scriptscriptstyle k}&\lambda_i\xi^{\scriptscriptstyle-i}+\lambda_j\xi^{\scriptscriptstyle-j}
        \end{pmatrix}\in{I_{_2}(1,\eta)}$\\
        \begin{align*}
         \begin{vmatrix}
             \lambda_i\xi^{\scriptscriptstyle i}+\lambda_j\xi^{\scriptscriptstyle j}&\lambda_k\xi^{\scriptscriptstyle-k}\\
            \lambda_k\xi^{\scriptscriptstyle k}&\lambda_i\xi^{\scriptscriptstyle-i}+\lambda_j\xi^{\scriptscriptstyle-j}
        \end{vmatrix}&=(\lambda_i\xi^{\scriptscriptstyle i}+\lambda_j\xi^{\scriptscriptstyle j})(\lambda_i\xi^{\scriptscriptstyle-i}+\lambda_j\xi^{\scriptscriptstyle-j})-\lambda_k^{\scriptscriptstyle2}=0\\
        &\Rightarrow(\lambda_i\xi^{\scriptscriptstyle i}+\lambda_j\xi^{\scriptscriptstyle j})(\lambda_i\xi^{\scriptscriptstyle-i}+\lambda_j\xi^{\scriptscriptstyle-j})=\lambda_k^{\scriptscriptstyle2}
    \end{align*}
\begin{equation*}
            \begin{pmatrix}
             \lambda_i\xi^{\scriptscriptstyle i}+\lambda_j\xi^{\scriptscriptstyle j}&\lambda_k\xi^{\scriptscriptstyle-k}\\
            \lambda_k\xi^{\scriptscriptstyle k}&\lambda_i\xi^{\scriptscriptstyle-i}+\lambda_j\xi^{\scriptscriptstyle-j}
        \end{pmatrix}=(\lambda_i\xi^{\scriptscriptstyle i}+\lambda_j\xi^{\scriptscriptstyle j})
        \begin{pmatrix}
            1&\dfrac{\lambda_k\xi^{\scriptscriptstyle-k}}{\lambda_i\xi^{\scriptscriptstyle i}+\lambda_j\xi^{\scriptscriptstyle j}}\\
            0&0
        \end{pmatrix}+(\lambda_k\xi^{\scriptscriptstyle k})
        \begin{pmatrix}
            0&0\\
            1&\dfrac{\lambda_i\xi^{\scriptscriptstyle-i}+\lambda_j\xi^{\scriptscriptstyle-j}}{\lambda_k\xi^{\scriptscriptstyle k}}
        \end{pmatrix}
\end{equation*}
Hence $\dfrac{\lambda_k\xi^{\scriptscriptstyle-k}}{\lambda_i\xi^{\scriptscriptstyle i}+\lambda_j\xi^{\scriptscriptstyle j}}=\dfrac{\lambda_i\xi^{\scriptscriptstyle-i}+\lambda_j\xi^{\scriptscriptstyle-j}}{\lambda_k\xi^{\scriptscriptstyle k}}=\eta$ . Bring them in $(\lambda_i\xi^{\scriptscriptstyle i}+\lambda_j\xi^{\scriptscriptstyle j})(\lambda_i\xi^{\scriptscriptstyle-i}+\lambda_j\xi^{\scriptscriptstyle-j})=\lambda_k^{\scriptscriptstyle2}$ , we \\
\quad\\
derive 
$
\left\{
\begin{array}{ll}
\eta\xi^{\scriptscriptstyle k}(\lambda_i\xi^{\scriptscriptstyle i}+\lambda_j\xi^{\scriptscriptstyle j})&=\lambda_k\\
\eta^{\scriptscriptstyle-1}\xi^{\scriptscriptstyle -k}(\lambda_i\xi^{\scriptscriptstyle-i}+\lambda_j\xi^{\scriptscriptstyle-j})&=\lambda_k
\end{array}
\right.
$ . 
Add the equation $\lambda_i+\lambda_j+\lambda_k=0$ to it form a\\
\quad\\
new system of equations as follow ,
\begin{equation}
\begin{cases}
    \eta\xi^{\scriptscriptstyle (i+k)}\lambda_i+\eta\xi^{\scriptscriptstyle (j+k)}\lambda_i-\lambda_k=0\\
    \eta^{\scriptscriptstyle-1}\xi^{\scriptscriptstyle -(i+k)}\lambda_i+\eta^{\scriptscriptstyle-1}\xi^{\scriptscriptstyle -(j+k)}\lambda_i-\lambda_k=0\\
    \lambda_i+\lambda_j+\lambda_k=0
\end{cases}
\end{equation}
In other words , there exist $\lambda_i,\lambda_j,\lambda_k\neq0$ satisfies (16) . We derive
\begin{align*}
    \begin{vmatrix}
        \eta\xi^{\scriptscriptstyle (i+k)}&\eta\xi^{\scriptscriptstyle (j+k)}&-1\\
        \eta^{\scriptscriptstyle-1}\xi^{\scriptscriptstyle -(i+k)}&\eta^{\scriptscriptstyle-1}\xi^{\scriptscriptstyle -(j+k)}&-1\\
        1&1&1
    \end{vmatrix}&=
    \begin{vmatrix}
        \eta\xi^{\scriptscriptstyle (i+k)}+1&\eta\xi^{\scriptscriptstyle (j+k)}+1&0\\
        \eta^{\scriptscriptstyle-1}\xi^{\scriptscriptstyle -(i+k)}+1&\eta^{\scriptscriptstyle-1}\xi^{\scriptscriptstyle -(j+k)}+1&0\\
        1&1&1
    \end{vmatrix}\\
    &=(\eta\xi^{\scriptscriptstyle (i+k)}+1)(\eta^{\scriptscriptstyle-1}\xi^{\scriptscriptstyle -(j+k)}+1)-
    (\eta\xi^{\scriptscriptstyle (j+k)}+1)(\eta^{\scriptscriptstyle-1}\xi^{\scriptscriptstyle -(i+k)}+1)\\
    &=0
\end{align*}
Namely , $(\eta\xi^{\scriptscriptstyle (i+k)}+1)(\eta^{\scriptscriptstyle-1}\xi^{\scriptscriptstyle -(j+k)}+1)=
    (\eta\xi^{\scriptscriptstyle (j+k)}+1)(\eta^{\scriptscriptstyle-1}\xi^{\scriptscriptstyle -(i+k)}+1)$ . We consider the factor $(\eta\xi^{\scriptscriptstyle (i+k)}+1)$ in the equation , $\eta$ and $\xi^{\scriptscriptstyle (i+k)}$ are $(q-1)-th$ primitive root of unity and $n-th$ root of unity respectively . Suppose $t|n$ is the order of $\xi^{\scriptscriptstyle (i+k)}$ , then the order of $\eta\xi^{\scriptscriptstyle (i+k)}$ is $\dfrac{(q-1)t}{(q-1,t)}=q-1$ , $\eta\xi^{\scriptscriptstyle (i+k)}\neq-1$ . Likewise , the other factor in the equation is non-zero .
\begin{align*}
&(\eta\xi^{\scriptscriptstyle (i+k)}+1)(\eta^{\scriptscriptstyle-1}\xi^{\scriptscriptstyle -(j+k)}+1)=(\eta\xi^{\scriptscriptstyle (j+k)}+1)(\eta^{\scriptscriptstyle-1}\xi^{\scriptscriptstyle -(i+k)}+1)\\
\Rightarrow&(\eta\xi^{\scriptscriptstyle (i+k)}+1)(\eta^{\scriptscriptstyle-1}\xi^{\scriptscriptstyle -(j+k)}+1)=\eta\xi^{\scriptscriptstyle(j+k)}\eta^{\scriptscriptstyle-1}\xi^{\scriptscriptstyle-(i+k)}(\eta\xi^{\scriptscriptstyle (i+k)}+1)(\eta^{\scriptscriptstyle-1}\xi^{\scriptscriptstyle -(j+k)}+1)\\
\Rightarrow&(\eta\xi^{\scriptscriptstyle (i+k)}+1)(\eta^{\scriptscriptstyle-1}\xi^{\scriptscriptstyle -(j+k)}+1)=\xi^{\scriptscriptstyle(j-i)}(\eta\xi^{\scriptscriptstyle (i+k)}+1)(\eta^{\scriptscriptstyle-1}\xi^{\scriptscriptstyle -(j+k)}+1)\\
\Rightarrow&1=\xi^{\scriptscriptstyle(j-i)}
\end{align*}
It is contradictory to $0\leq i<j\leq n-1$ and $\xi$ is $n-th$ primitive of unity .\\
\quad\\
case 3: If  $\lambda_{i}a^{\scriptscriptstyle i}+\lambda_{j}ba^{\scriptscriptstyle j}+\lambda_{k}ba^{\scriptscriptstyle k}\:,\:0\leq j<k\leq n-1\text{ and } \lambda_i,\lambda_j,\lambda_k\neq0$ . 
\begin{align*}
    P(u)=\begin{pmatrix}
          \lambda_i+\lambda_j+\lambda_k\quad,&\lambda_i-\lambda_j-\lambda_k
        \end{pmatrix}
        \times
        \begin{pmatrix}
             \lambda_i\xi^{\scriptscriptstyle i}&\lambda_j\xi^{\scriptscriptstyle-j}+\lambda_k\xi^{\scriptscriptstyle-k}\\
            \lambda_j\xi^{\scriptscriptstyle j}+\lambda_k\xi^{\scriptscriptstyle k}&\lambda_i\xi^{\scriptscriptstyle-i}
        \end{pmatrix}
        \times\cdots\times
\end{align*}
Then ,\\
    $\begin{pmatrix}
           \lambda_i+\lambda_j+\lambda_k\quad,&\lambda_i-\lambda_j-\lambda_k 
        \end{pmatrix}\in{I_{_1}(0,1)}
        \:,\:
        \begin{pmatrix}
            \lambda_i\xi^{\scriptscriptstyle i}&\lambda_j\xi^{\scriptscriptstyle-j}+\lambda_k\xi^{\scriptscriptstyle-k}\\
            \lambda_j\xi^{\scriptscriptstyle j}+\lambda_k\xi^{\scriptscriptstyle k}&\lambda_i\xi^{\scriptscriptstyle-i}
        \end{pmatrix}\in{I_{_2}(1,\eta)}$\\
        \begin{align*}
         \begin{vmatrix}
            \lambda_i\xi^{\scriptscriptstyle i}&\lambda_j\xi^{\scriptscriptstyle-j}+\lambda_k\xi^{\scriptscriptstyle-k}\\
            \lambda_j\xi^{\scriptscriptstyle j}+\lambda_k\xi^{\scriptscriptstyle k}&\lambda_i\xi^{\scriptscriptstyle-i}
        \end{vmatrix}&=\lambda_i^{\scriptscriptstyle2}-(\lambda_j\xi^{\scriptscriptstyle j}+\lambda_k\xi^{\scriptscriptstyle k})(\lambda_j\xi^{\scriptscriptstyle-j}+\lambda_k\xi^{\scriptscriptstyle-k})=0\\
        &\Rightarrow(\lambda_j\xi^{\scriptscriptstyle j}+\lambda_k\xi^{\scriptscriptstyle k})(\lambda_j\xi^{\scriptscriptstyle-j}+\lambda_k\xi^{\scriptscriptstyle-k})=\lambda_k^{\scriptscriptstyle2}
\end{align*}
\begin{equation*}
        \begin{pmatrix}
            \lambda_i\xi^{\scriptscriptstyle i}&\lambda_j\xi^{\scriptscriptstyle-j}+\lambda_k\xi^{\scriptscriptstyle-k}\\
            \lambda_j\xi^{\scriptscriptstyle j}+\lambda_k\xi^{\scriptscriptstyle k}&\lambda_i\xi^{\scriptscriptstyle-i}
        \end{pmatrix}=(\lambda_i\xi^{\scriptscriptstyle i})
        \begin{pmatrix}
            1&\dfrac{\lambda_j\xi^{\scriptscriptstyle-j}+\lambda_k\xi^{\scriptscriptstyle-k}}{\lambda_i\xi^{\scriptscriptstyle i}}\\
            0&0
        \end{pmatrix}+(\lambda_j\xi^{\scriptscriptstyle j}+\lambda_k\xi^{\scriptscriptstyle k})
        \begin{pmatrix}
            0&0\\
            1&\dfrac{\lambda_i\xi^{\scriptscriptstyle-i}}{\lambda_j\xi^{\scriptscriptstyle j}+\lambda_k\xi^{\scriptscriptstyle k}}
        \end{pmatrix}
\end{equation*}
Hence $\dfrac{\lambda_j\xi^{\scriptscriptstyle-j}+\lambda_k\xi^{\scriptscriptstyle-k}}{\lambda_i\xi^{\scriptscriptstyle i}}=\dfrac{\lambda_i\xi^{\scriptscriptstyle-i}}{\lambda_j\xi^{\scriptscriptstyle j}+\lambda_k\xi^{\scriptscriptstyle k}}=\eta$ . Bring them in $(\lambda_j\xi^{\scriptscriptstyle j}+\lambda_k\xi^{\scriptscriptstyle k})(\lambda_j\xi^{\scriptscriptstyle-j}+\lambda_k\xi^{\scriptscriptstyle-k})=\lambda_k^{\scriptscriptstyle2}$ ,\\
\quad\\
we derive 
$
\left\{
\begin{array}{ll}
\eta\xi^{\scriptscriptstyle i}(\lambda_j\xi^{\scriptscriptstyle j}+\lambda_k\xi^{\scriptscriptstyle k})&=\lambda_i\\
\eta^{\scriptscriptstyle-1}\xi^{\scriptscriptstyle -i}(\lambda_j\xi^{\scriptscriptstyle-j}+\lambda_k\xi^{\scriptscriptstyle-k})&=\lambda_i
\end{array}
\right.
$ . 
Add the equation $\lambda_i+\lambda_j+\lambda_k=0$ to it form a\\
\quad\\
new system of equations as follow ,
\begin{equation}
\begin{cases}
    \eta\xi^{\scriptscriptstyle (i+j)}\lambda_j+\eta\xi^{\scriptscriptstyle (i+k)}\lambda_k-\lambda_i=0\\
    \eta^{\scriptscriptstyle-1}\xi^{\scriptscriptstyle -(i+j)}\lambda_j+\eta^{\scriptscriptstyle-1}\xi^{\scriptscriptstyle -(i+k)}\lambda_k-\lambda_i=0\\
    \lambda_i+\lambda_j+\lambda_k=0
\end{cases}
\end{equation}
In other words , there exist $\lambda_i,\lambda_j,\lambda_k\neq0$ satisfies (16) . We derive
\begin{align*}
    \begin{vmatrix}
        \eta\xi^{\scriptscriptstyle (i+j)}&\eta\xi^{\scriptscriptstyle (i+k)}&-1\\
        \eta^{\scriptscriptstyle-1}\xi^{\scriptscriptstyle -(i+j)}&\eta^{\scriptscriptstyle-1}\xi^{\scriptscriptstyle -(i+k)}&-1\\
        1&1&1
    \end{vmatrix}&=
    \begin{vmatrix}
        \eta\xi^{\scriptscriptstyle (i+j)}+1&\eta\xi^{\scriptscriptstyle (i+k)}+1&0\\
        \eta^{\scriptscriptstyle-1}\xi^{\scriptscriptstyle -(i+j)}+1&\eta^{\scriptscriptstyle-1}\xi^{\scriptscriptstyle -(i+k)}+1&0\\
        1&1&1
    \end{vmatrix}\\
    &=(\eta\xi^{\scriptscriptstyle (i+j)}+1)(\eta^{\scriptscriptstyle-1}\xi^{\scriptscriptstyle -(i+k)}+1)-
    (\eta\xi^{\scriptscriptstyle (i+k)}+1)(\eta^{\scriptscriptstyle-1}\xi^{\scriptscriptstyle -(i+j)}+1)\\
    &=0
\end{align*}
\begin{align*}
&(\eta\xi^{\scriptscriptstyle (i+j)}+1)(\eta^{\scriptscriptstyle-1}\xi^{\scriptscriptstyle -(i+k)}+1)=(\eta\xi^{\scriptscriptstyle (i+k)}+1)(\eta^{\scriptscriptstyle-1}\xi^{\scriptscriptstyle -(i+j)}+1)\\
\Rightarrow&(\eta\xi^{\scriptscriptstyle (i+j)}+1)(\eta^{\scriptscriptstyle-1}\xi^{\scriptscriptstyle -(i+k)}+1)=\eta\xi^{\scriptscriptstyle (i+k)}\eta^{\scriptscriptstyle-1}\xi^{\scriptscriptstyle -(i+j)}(\eta^{\scriptscriptstyle-1}\xi^{\scriptscriptstyle -(i+k)}+1)(\eta\xi^{\scriptscriptstyle (i+j)}+1)\\
\Rightarrow&(\eta\xi^{\scriptscriptstyle (i+j)}+1)(\eta^{\scriptscriptstyle-1}\xi^{\scriptscriptstyle -(i+k)}+1)=\xi^{\scriptscriptstyle(k-j)}(\eta^{\scriptscriptstyle-1}\xi^{\scriptscriptstyle -(i+k)}+1)(\eta\xi^{\scriptscriptstyle (i+j)}+1)\\
\Rightarrow&1=\xi^{\scriptscriptstyle(k-j)}
\end{align*}
It is contradictory to $0\leq j<k\leq n-1$ and $\xi$ is $n-th$ primitive of unity .\\
\quad\\
case 4: It is similar to case 1 .
\end{proof}
Next we replace the condition $2n<q-1$ by $2n\leq q-1$ and give another MDS dihedral group code .\\
\textbf{Corollary 4.7}. Let $I=\mathop{\bigoplus}\limits_{j\neq 0,1,n-1}Re_{_j}\oplus R(\dfrac{1-b}{2}e_{_0})\oplus R(e_{_1}+\eta be_{_{n-1}})$ , The generate matrix of $phi(I)$ is 
\begin{equation*}
    G_{I}=\begin{pmatrix}
        \phi(\frac{1-b}{2}e_{_0})\\
        \phi(e_{_1}+\eta be_{_{n-1}})\\
        \phi(be_{_1}+\eta e_{_{n-1}})\\
        \phi(e_{_2})\\
        \phi(be_{_2})\\
        \vdots\\
        \phi(e_{_{n-2}})\\
        \phi(be_{_{n-2}})
    \end{pmatrix}
    =\begin{pmatrix}
        1&1&\cdots&1&-1&-1&\cdots&-1\\
        1&\xi^{\scriptscriptstyle-1}&\cdots&\xi^{\scriptscriptstyle-(n-1)}&\eta&\eta\xi^{\scriptscriptstyle1}&\cdots&\eta\xi^{\scriptscriptstyle n-1}\\
        \eta&\eta\xi^{\scriptscriptstyle1}&\cdots&\eta\xi^{\scriptscriptstyle n-1}&1&\xi^{\scriptscriptstyle-1}&\cdots&\xi^{\scriptscriptstyle-(n-1)}\\
        1&\xi^{\scriptscriptstyle-2}&\cdots&\xi^{\scriptscriptstyle-2(n-1)}&0&0&\cdots&0\\
        0&0&\cdots&0&1&\xi^{\scriptscriptstyle-2}&\cdots&\xi^{\scriptscriptstyle-2(n-1)}\\
        \vdots&\vdots&\quad&\vdots&\vdots&\vdots&\quad&\vdots\\
        1&\xi^{\scriptscriptstyle-(n-2))}&\cdots&\xi^{\scriptscriptstyle-(n-2)(n-1)}&0&0&\cdots&0\\
        0&0&\cdots&0&1&\xi^{\scriptscriptstyle-(n-2)}&\cdots&\xi^{\scriptscriptstyle-(n-2)(n-1)}
    \end{pmatrix}
\end{equation*}
\textbf{Corollary 4.8}. Let $I=\mathop{\bigoplus}\limits_{j\neq 0,1,n-1}Re_{_j}\oplus R(\dfrac{1+b}{2}e_{_0})\oplus R(e_{_1}+\eta be_{_{n-1}})$ is a left ideal of $R$ , then $dim_{_{F_q}}I=2n-3$ , $\phi(I)$ is $[2n,2n-3,4]$ MDS dihedral code .
\begin{proof}
    Use the notation above in \textbf{Lemma 4.1} , we derive
\[
P(I)=\bigoplus\limits_{j=1}^{1+\frac{n-1}{2}}B_{_j},\qquad 
    B_{_j}=\begin{cases}
    A_{_j}&,j\in{\{3,4\cdots,1+\dfrac{n-1}{2}\}}\\
    I_{_j}(1,0)&,j=1\\
    I_{_j}(1,\eta)&,j=2
\end{cases}
\]
From the prove of \textbf{Theorem 4.2} and \textbf{Theorem 4.6} , the cases are similar except the case that $u=\lambda_ia^{\scriptscriptstyle i}+\lambda_jba^{\scriptscriptstyle j}\notin{I}$ , for any $\lambda_i,\lambda_j$ and $0\leq i,j\leq n-1$ . Thus we only need to verify this case . By the case 2 of \textbf{Theorem 4.2} , 
$\begin{pmatrix}
    \lambda_i+\lambda_j&\lambda_i-\lambda_j
\end{pmatrix}\in{I_{_2}}(1,0)$ , 
$\begin{pmatrix}
    \lambda_i\xi^{\scriptscriptstyle i}&\lambda_j\xi^{\scriptscriptstyle-j}\\
    \lambda_j\xi^{\scriptscriptstyle j}&\lambda_i\xi^{\scriptscriptstyle-i}
\end{pmatrix}\in{I_{_2}}(1,\eta)$ . And
\begin{equation*}
    \begin{pmatrix}
      \lambda_i\xi^{\scriptscriptstyle i}&\lambda_j\xi^{\scriptscriptstyle-j}\\
      \lambda_j\xi^{\scriptscriptstyle j}&\lambda_i\xi^{\scriptscriptstyle-i}
    \end{pmatrix}=(\lambda_i\xi^{\scriptscriptstyle i})
        \begin{pmatrix}
            1&\dfrac{\lambda_j\xi^{\scriptscriptstyle-j}}{\lambda_i\xi^{\scriptscriptstyle i}}\\
            0&0
        \end{pmatrix}+(\lambda_j\xi^{\scriptscriptstyle j})
        \begin{pmatrix}
            0&0\\
            1&\dfrac{\lambda_i\xi^{\scriptscriptstyle-i}}{\lambda_j\xi^{\scriptscriptstyle j}}
        \end{pmatrix}\in{I_{_2}}(1,\eta)
\end{equation*}
Namely , $\dfrac{\lambda_j\xi^{\scriptscriptstyle-j}}{\lambda_i\xi^{\scriptscriptstyle i}}=\dfrac{\lambda_i\xi^{\scriptscriptstyle-i}}{\lambda_j\xi^{\scriptscriptstyle j}}=\eta$ , $\lambda_i-\lambda_j=0$ . Then we can derive there exist $\lambda_i,\lambda_j\neq0$ is the solve if the system of equations as follow , 
\begin{equation}
\begin{cases}
    \eta\xi^{\scriptscriptstyle i}\lambda_i-\xi^{\scriptscriptstyle -j}\lambda_j=0\\
    \xi^{\scriptscriptstyle -i}\lambda_i-\eta\xi^{\scriptscriptstyle j}\lambda_j=0\\
    \lambda_i-\lambda_j=0
\end{cases}
\end{equation}
We only consider the second and third equation of the system of equations , the determinant derived from them is 
$\begin{vmatrix}
    \xi^{\scriptscriptstyle -i}&-\eta\xi^{\scriptscriptstyle j}\\
    1&-1
\end{vmatrix}=\eta\xi^{\scriptscriptstyle j}-\xi^{\scriptscriptstyle -i}$ . However , $\lambda_i,\lambda_j\neq0$ implies that the determinant must be $0$ , that is $\eta\xi^{\scriptscriptstyle j}=\xi^{\scriptscriptstyle -i}\Rightarrow\eta=\xi^{\scriptscriptstyle -(i+j)}$ . It is contradictory to the condition that $\eta$ is $(q-1)-th$ primitive root of unity , $\xi^{\scriptscriptstyle -(i+j)}$ is $n-th$ root of unity and $2n\leq q-1$ .
\end{proof}
\textbf{Corollary 4.9}. If $\beta\in{F_q}$ and $\beta^{\scriptscriptstyle n}\neq1$ , let $I=\mathop{\bigoplus}\limits_{j\neq 0,1,n-1}Re_{_j}\oplus R(\dfrac{1+b}{2}e_{_0})\oplus R(e_{_1}+\beta be_{_{n-1}})$ is a left ideal of $R$ , then $\phi(I)$ is $[2n,2n-3,4]$ MDS dihedral code .
\begin{proof}
    It follows from the prove of \textbf{Theorem 4.2} , \textbf{Theorem 4.6} and \textbf{Corollary 4.9} .
\end{proof}
\section{Example}
\textbf{Example 5.1}. Let $F_{5^{\scriptscriptstyle2}}=F_5[x]/<x^{\scriptscriptstyle2}+1>$ be a finite field of order 25 , $D_{2\times3}$ is a dihedral group of order 6 . Since $6|24$ and $\eta=x+1$ is a $24-th$ primitive root of unity , then $\xi=\eta^8=x+2$ is a $3-th$ primitive root of unity . From \textbf{Corollary 2.2} , 
    \begin{align*}
        P:F_qD_{2\times3}&\longrightarrow\qquad\mathop{\bigoplus}\limits_{j=1}^{2}A_j\\
        a\quad&\longmapsto\begin{pmatrix}
            1&1
        \end{pmatrix}
        \times
        \begin{pmatrix}
            \xi&0\\
            0&\xi^{\scriptscriptstyle-1}
        \end{pmatrix}\\
        b\quad&\longmapsto\begin{pmatrix}
            1&-1
        \end{pmatrix}
        \times
        \begin{pmatrix}
            0&1\\
            1&0
        \end{pmatrix}
    \end{align*}
From \textbf{Corollary 3.5} , 
\begin{align*}
    &e_{_0}=\dfrac{1}{n}\sum\limits_{j=0}^{2}a^{\scriptscriptstyle i}=2+2a+2a^{\scriptscriptstyle2},\\
    &e_{_1}=\dfrac{1}{n}\sum\limits_{j=0}^{2}\xi^{\scriptscriptstyle-j}a^{\scriptscriptstyle i}=2+(3x+4)a+(2x+4)a^{\scriptscriptstyle2},\\
    &e_{_2}=\dfrac{1}{n}\sum\limits_{j=0}^{2}\xi^{\scriptscriptstyle-2j}a^{\scriptscriptstyle i}=2+(2x+4)a+(3x+4)a^{\scriptscriptstyle2}\\
    &\eta e_{_2}=\dfrac{1}{n}\eta\sum\limits_{j=0}^{2}\xi^{\scriptscriptstyle-2j}a^{\scriptscriptstyle i}=(2x+2)+xa+(2x+3)a^{\scriptscriptstyle2}
\end{align*}
Let $I_{_1}=Re_{_0}\oplus R(e_{_1}+\eta be{_2})$ , then the generator matrix of $\phi(I_{_1})$ is 
\begin{equation*}
    G_{I_{_1}}=\begin{pmatrix}
        1&1&1&0&0&0\\
        0&0&0&1&1&1\\
        2&3x+4&2x+4&2x+2&x&2x+3\\
        2x+2&x&2x+3&2&3x+4&2x+4
    \end{pmatrix}
\end{equation*}
We use \textbf{Sagemath} to calculate the minimum distance of $\phi(I_{_1})$ and $d(\phi(I_{_1}))=3$ . We can also verify that any $4\times4$ submatrix of $G_{I_{_1}}$ is non-degenerate to derive the minimum distance of $\phi(I_{_1})$ is $3$ . Thus $\phi(I_{_1})$ is $[6,4,3]$ MDS dihedral group code .\\
Let $I_{_2}=R(\dfrac{1+b}{2}e_{_0})\oplus R(e_{_1}+\eta be{_2})$ , then the generator matrix of $\phi(I_{_2})$ is 
\begin{equation*}
    G_{I_{_1}}=\begin{pmatrix}
        1&1&1&1&1&1\\
        2&3x+4&2x+4&2x+2&x&2x+3\\
        2x+2&x&2x+3&2&3x+4&2x+4
    \end{pmatrix}
\end{equation*}
We use \textbf{Sagemath} to calculate the minimum distance of $\phi(I_{_2})$ and $d(\phi(I_{_2}))=4$ . We can also verify that any $3\times3$ submatrix of $G_{I_{_2}}$ is non-degenerate to derive the minimum distance of $\phi(I_{_2})$ is $4$ . Thus $\phi(I_{_1})$ is $[6,3,4]$ MDS dihedral group code .\\

\newpage


\bibliographystyle{unsrt}
\bibliography{reference}

\end{document}